\documentclass[letterpaper,10pt]{article}

\twocolumn

\usepackage[latin1]{inputenc}
\usepackage{graphicx,float,xspace,latexsym}

\newcommand{\Iac}{$I_{AC}$\@\xspace}
\newcommand{\Idc}{$I_{DC}$\@\xspace}
\newcommand{\Imm}{$I_{M}$\@\xspace}
\newcommand{\Ibias}{$I_{bias}$\@\xspace}
\newcommand{\Imin}{$I_{min}$\@\xspace}
\newcommand{\Imax}{$I_{MAX}$\@\xspace}
\newcommand{\Ic}{$I_{c}$\@\xspace}

\begin{document}
\newpage
\pagestyle{empty}
\thispagestyle{empty}
\title{Effect of Combining a DC Bias Current with an AC Transport Current on 
AC Losses in a High Temperature Superconductor}

\date{}

\author{Patricia Dolez, Benoît des Ligneris, Marcel Aubin \\{\footnotesize 
Département de Physique and Centre de Recherche en Physique du Solide -- 
Université de Sherbrooke -- Québec -- Canada} \\ Wen Zhu, Julian Cave \\ 
{\footnotesize Technologies émergentes de production et de stockage -- VPTI 
Hydro-Québec -- Varennes -- Québec -- Canada}\\ \vspace{.2cm}{\footnotesize 
This work has been financed jointly by the IREQ (Hydro-Québec) and the 
NSERC.}}

\vspace{-1in}

\maketitle
\thispagestyle{empty}

\begin{abstract}
Creating complex flux configurations by superposing a dc current or magnetic 
field onto the ac current in a type II superconducting tape should lead to a 
variety of peculiar behaviors. An example is the appearance of the Clem 
valley, a minimum in the ac losses as a function of the dc bias amplitude, 
which has been theoretically studied by LeBlanc et al., in the continuation 
of Clem's calculations. These situations have been investigated by applying 
a dc current to a silver-gold sheathed Bi-2223 tape at 77~K (critical 
current 29~A), in addition to the usual ac transport current. The ac losses 
were measured by the null calorimetric method to ensure that the total 
losses were being accounted for. These were recorded for different values of 
the ac and dc currents, leading to the observation of two different 
behaviors depending on the ac current. Our revelation of the Clem valley is, 
to our knowledge, the first experimental validation of this phenomenon in 
high temperature superconductors, and may provide a simple way of reducing 
the ac loss in industrial applications of these materials.
\end{abstract}

\section{\sc Introduction}
Since their discovery, high temperature superconductors (HTS) have seen 
their properties greatly improved following the optimization of their 
fabrication process and their design. Their widespread use in industrial 
applications seems to be rapidly approaching. Nevertheless, the study of 
their behavior has generally included very simple arrangements, whereas 
practical situations will involve much more complex situations, for example 
the electromagnetic environment.

In the particular field of energy transportation, HTS tapes may be subjected 
not only to an ac current, but also to magnetic fields of various 
orientations from neighboring conductor tapes, or from other devices such as 
coils in a transformer. Thus, while the isolated superconducting tape 
behavior in transport current conditions seems to be quite well understood 
by way of intensive ac loss measurements, interest now turns towards 
subjecting the superconducting tape to the superposition of several 
electromagnetic perturbations. 

For example, some workers are measuring the effect of dc 
\cite{Ciszek96,Dutoit97,Benni98} and ac \cite{Benni98,Fukui96,Daney97,Mag97} 
magnetic fields on ac transport losses or the effect of dc current on ac 
losses induced by an ac magnetic field \cite{Ciszek97}.

Following previous work of Clem \cite{Clem79} and later by LeBlanc et al. 
\cite{Leblanc86}, we choose to apply a dc current to our Ag-Au/Bi-2223 
superconducting tape in addition to the usual ac transport one, and to look 
at the effects on its ac losses. This paper presents our first results, 
which confirm the expected decrease in ac losses, and the suggested 
explanations for the observed effects.

\section{\sc Theoretical considerations}

When the ac losses in a tape are generated by the application of an ac 
transport current or an ac magnetic field, the profiles of magnetic 
induction inside the superconducting materials show a horizontal mirror 
symmetry, which leads to the well-known ac loss calculations. On the other 
hand, if a dc signal (collinear current or magnetic field) is applied to the 
tape in addition to the ac one, flux density profiles may lose this symmetry 
if one considers the field dependence of the bulk pinning properties 
\cite{Leblanc86,Leblanc96} (by way of the critical current), or the presence 
of surface barriers against flux entry and exit \cite{Clem79,Leblanc95} or 
the Meissner screening current \Imm\cite{Leblanc96,Leblanc95}.

\begin{figure}[h]
\centering
\includegraphics[width=3in]{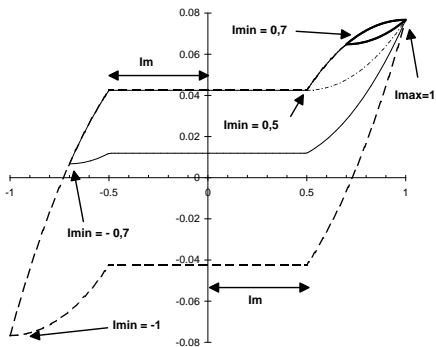}
\caption{\footnotesize Schematic of the hysteresis loops of the magnetic 
flux density while the total current is varied between \Imax and -\Imin 
(\Imax~=~\Iac~+~\Ibias, and \Imin~=~\Ibias~-~\Iac), taking into account a 
Meissner current \Imm. For the displayed family of hysteresis loops, Imax is 
kept constant, and \Ibias is increased as \Iac is decreased.}
\label{Hysteresis}
\end{figure}

We consider in particular the latter which is a current circulating at the 
surface to compensate the magnetic field outside the superconducting 
material when in the Meissner state, and which modifies the hysteresis loops 
of the magnetic flux density in such a way that the total current flowing 
through the material is reduced by \Imm in absolute value. This situation is 
illustrated in Fig.\@\xspace\ref{Hysteresis}, using the theoretical 
formulation established by LeBlanc et al. \cite{Leblanc96}, in the case 
where a dc current \Ibias is applied in addition to the ac transport current 
\Iac. The average magnetic flux density is given as a function of the total 
current, which is being varied between \{\Imax~=~\Iac~+~\Ibias\} and 
\{\Imin~=~\Ibias~-~\Iac~\}. Here the critical current is considered 
independent of the magnetic field, following the Bean approximation.

When expressing the ac losses as a function of the applied dc bias current, 
two different behaviors are observed depending on whether the ac current is 
smaller or larger than \Imm. 

\begin{figure}[h]
\centering
\includegraphics[width=3in]{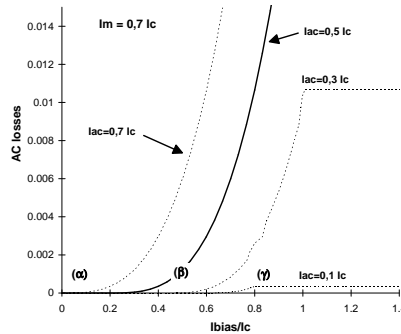}
\caption{\footnotesize Graph of the theoretical calculation of the ac losses 
as a function the dc bias current, with a 0.7Ic Meissner screening current 
and in the elliptical geometry, for four values of the ac current less than 
the Meissner screening current (0.1, 0.3, 0.5 and 0.7~\Ic).}
\label{Theorieplateau}
\end{figure}

If the ac current amplitude is less than the Meissner screening current, 
then the ac losses have the shape displayed in 
Fig.\@\xspace\ref{Theorieplateau}. These ac losses curves are also obtained 
using the LeBlanc et al. calculations \cite{Leblanc96}, with each one 
corresponding to a value of the ac current and all current values normalized 
by the critical current \Ic. The initial plateau (identified as zone 
$\alpha$) appears when the total current (\Iac + \Idc) is less than the 
Meissner current, i.e. when the magnetic flux is totally expelled from the 
superconductor: there are no losses at all. Then the ac losses increase 
(zone $\beta$) as the dc current is increased since screening is no longer 
totally effective. A last zone $\gamma$, with another plateau, may appear as 
the dc current is increased above \{\Iac~+~\Imm\} if the condition 
\{\Iac~$<$~\Ic/2\} is met. Otherwise, the ac losses variation ends with the 
($\beta$) zone. The $\gamma$ zone plateau occurs when \Imin~=~\{\Ibias~-
~\Iac\} is larger than the Meissner current, so that the hysteresis loop 
displays no horizontal segment. This case is illustrated by the thick curves 
in Fig.\@\xspace\ref{Theorieplateau}.

If the ac current is greater than the Meissner screening current, the ac 
loss behavior as a function of the dc bias current is different from the 
previous situation. In that case, the first plateau (zone $\alpha$) is 
replaced by a decrease, so that the ac loss curves display a minimum as the 
dc current amplitude is augmented. On the other hand, the two other 
typicallyshaped zones ($\beta$ and $\gamma$) still exist and follow the same 
criteria. The reason for this decrease in ac losses is the diminution of the 
area between the two horizontal segments of the hysteresis loops (see 
Fig.\@\xspace\ref{Theorieplateau}) as the dc current amplitude is augmented. 
The location of the ac loss minimum can be deduced from the 
Fig.\@\xspace\ref{Theorieplateau} curves and corresponds to a superposition 
of the upper and lower horizontal segments, i.e. to \Imin~=~-\Imm. Thus the 
minimum occurs when \Ibias~=~\Iac~-~\Imm.

A similar behavior for the ac loss as a function of the dc bias current is 
obtained when replacing the Bean approximation by the Kim approximation, 
i.e. with the critical current inversely proportional to the magnetic 
induction. The only difference consists in more rounded transitions between 
the different zones.

\section{Experimental details}

The null calorimetric method \cite{These_pat,Pat98} has been used for the ac 
loss measurements at liquid nitrogen temperature. Very briefly, its 
principle consists in feeding alternately during a few seconds the 
superconducting tape with an ac current (plus a dc bias current in this 
case) and a reference tape with an increasing dc current, until their 
heating effects are equal so that the amplitude of the temperature variation 
is nulled. The major advantage of a calorimetric method is the assurance of 
recording all the losses, which becomes fundamental when dealing with more 
complex electromagnetic environments. For example, this is not the case for 
the electrical method, where one is restricted to only the transport loss.

The superconductor tape measured here is a silver-gold alloy sheathed Bi-
2223 conductor made by the Powder-in-Tube technique in IREQ laboratories. 
Its characteristic dimensions are a cross-sectional area of 2.10$^{-3}$ 
cm$^2$, and a length of about 30 cm. Its critical current has been 
determined by the fitting of its V-I characteristic with the double 
integration of the sum of gaussian curves and is equal to 29 A. Excessive 
heating occurred during the measurement campaign and damaged the sample 
thereby lowering its critical current value to about 25 A. The data 
corresponding to one or the other of the sample conditions, i.e. to a 29 or 
25 A critical current, will be clearly identified in 
Fig.\@\xspace\ref{Theorieplateau} and Fig.\@\xspace\ref{Expplateau}, which 
illustrate the measurement results.

The reference tape mentioned above is composed of brass (resistivity of 7.3 
$\mu\Omega$.cm at 77 K), and has the same dimensions as the superconducting 
sample.

\section{Results and discussion}

The ac losses were measured for different values of the ac and dc currents 
in the superconducting tape. The choice of the ac current frequency was 
guided by the requirement of a maximum resolution in the determination of 
the ac losses, but without heating the sample excessively. Consequently, the 
ac loss measurements at lower currents (less than half the critical current) 
were made at 559 Hz, while those at higher currents were done at 55 Hz. 
Nevertheless, we verified that the choice of the frequency has no effect on 
the shape of the ac loss curve.

When expressed as a function of the dc bias current, the measured ac losses 
display two different behaviors depending on the value of the fixed ac 
current.

\begin{figure}[h]
\centering
\includegraphics[width=3in]{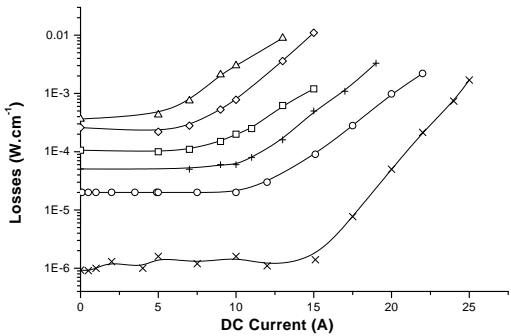}
\caption{\footnotesize Graph of the measured ac losses as a function of the 
dc current for six values of the ac transport current (at 559 Hz), ($\times$ 
for 3 Arms, o for 7, + for 9, $\Box$ for 11, $\Diamond$ for 13 and 
$\triangle$ for 15). The critical current is 29.~A.}
\label{Expplateau}
\end{figure}

\begin{figure}[h]
\centering
\includegraphics[width=3in]{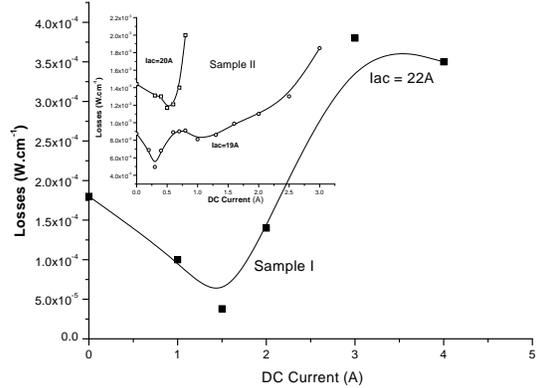}
\caption{\footnotesize Graphes of the measured ac losses as a function of 
the dc current for three values of the ac transport current (at 55 Hz). The 
22 Arms measurement deals with a sample critical current of 29 A, while for 
the 19 and 20 Arms ones in inset), the critical current is lowered to about 
25 A.}
\label{Expvallee}
\end{figure}

As shown in Fig.\@\xspace\ref{Expplateau}, for low ac current values 
(ranging from 3 to 15 Arms, while the critical current is 29 A), the ac loss 
curves as a function of the dc bias current show an initial plateau at a 
loss level equal to that obtained with no additional dc current, followed by 
an increase. In this graph, ac losses, expressed in W/cm, are measured at 
559 Hz and displayed on a log scale for convenience. Solid and dotted lines 
through the data points serve as guides to the eye. The dc current values 
corresponding to the end of the plateau decrease as the ac current is 
augmented, and seem to follow the following relationship: 
Idc~+~Iac~=~constant.

For the higher ac current values, i.e. 19 and 20~Arms with a critical 
current value of 25~A, and 22~Arms for a critical current of 29~A, the ac 
loss curves in Fig.\@\xspace\ref{Expvallee} display an initial decrease as 
the dc current value is increased, followed by the usual increase. When 
comparing the 19 and 20~Arms curves, one can see that the location of this 
minimum moves to higher dc current values as the ac current amplitude is 
increased, although corresponding dc current values remain very small (less 
than 1~A). A more thorough investigation is required to characterize this 
movement. It appears that these are the first reported observations of the 
Clem valley in HTS and the first in any superconductor using transport 
currents. LeBlanc et al. had observed it on a VTi sample with applied 
magnetic fields \cite{Leblanc86}. 

Returning to the loss behavior at low ac currents, our measurements display 
the general shape of LeBlanc's theoretical predictions (see 
Fig.\@\xspace\ref{Theorieplateau}), with a plateau followed by an ac loss 
increase. Moreover, the position of the end of the plateau seems to obey the 
predicted relationship, i.e. \Idc~+~\Iac~=~constant. In the simulations, 
this constant, 20 A from our measurements, would correspond to the Meissner 
screening current. 

Nevertheless, the loss level of the plateau ($\alpha$ zone) constitutes a 
major discrepancy, since the Meissner screening current is expected to 
prevent any flux line from entering the superconducting material whenever 
the total applied current (\Iac~+~\Idc) is less than \Imm, and consequently, 
the plateau is situated at a zero ac loss level. On the contrary, in our 
measurements, the plateau height for each ac current value corresponds to 
the ac losses measured when no additional dc current is applied, which are 
anything but null. Moreover, for an HTS material, the Meissner screening 
current value, which is related to the lowest critical magnetic field, is 
very small (of the order of magnitude of~10$^{-5}$A).

A possible explanation for this discrepancy comes from the granular nature 
of our sample. Before penetrating into the superconducting grains, magnetic 
flux lines first enter the intergrain material, which possesses a lower 
critical current density and which acts as weak links between the grains. 
Therefore, each grain is individually screened from the magnetic flux lines 
by the intergrain material, and the current circulating in it. In this 
regime, the losses are solely due to the ac current in the intergrain 
superconducting material. When an increase in the dc bias current causes the 
total current to exceed the intergrain critical current, flux lines begin to 
enter into the grains, which then also contribute to the ac losses. Any 
further increase of the dc current allows more flux lines to penetrate into 
the grains, which leads to an increase in the losses. According to this 
interpretation, the intergrain critical current is estimated to be equal to 
20~A, while the intragrain one is 29~A (these measurements at low ac 
currents have been made on the sample before its deterioration).

\section{Conclusion}

The measurement of ac losses on a Ag-Au/Bi-2223 superconducting tape has 
been performed with a dc bias current superimposed on the ac transport 
current. The results are generally coherent with the simulations of LeBlanc 
et al. but a discrepancy remains in the loss level at low ac currents. While 
more theoretical and experimental work is required in this matter, we may 
point to the existence of both intragrain and intergrain material in real 
samples. Nevertheless, our results constitute the first observation of the 
Clem valley in HTS and the first in any superconductor by a transport 
method.

\section*{Acknowledgment}

The authors want to thank Professor M.A.R. LeBlanc for his kind help.

\end{document}